\def\gapprox{{_>\atop{^\sim}}}
\def\lapprox{{_<\atop{^\sim}}}

\def\cmmd{\rm {cm^{-3}}}

\def\s-1{\rm {s^{-1}}}

\def\etal {et al.}
\def\kms {\hbox{${\rm km\,s}^{-1}$}}
\documentclass{aa}
\usepackage{graphicx}
\begin{document}
\title{CN and HNC Line Emission in IR Luminous Galaxies}
\author{S.~Aalto\inst{1},  A. G. Polatidis\inst{1}, S.~H\"uttemeister\inst{1,2},
S. J. Curran\inst{3}}
\offprints{S. Aalto}
\institute{
 Onsala Rymdobservatorium, Chalmers Tekniska H\"ogskola, S - 439 92 Onsala, Sweden
\and
Astronomisches Institut der Universit\"{a}t Bochum,
 Universit\"atsstra\ss{}e 150,  D - 44780 Bochum, Germany
 \and
 School of Physics,
 University of New South Wales,
  Sydney NSW 2052,
  Australia}
\date{Received  / Accepted  }
\titlerunning{CN,HNC in Galaxies}

%
\abstract{We have observed HNC 1-0, CN 1-0 \& 2-1 line emission in a sample of 13 IR
luminous (LIRGs, $L_{\rm IR} > 10^{11}$ L$_{\odot}$) starburst and Seyfert galaxies. 
HNC 1-0 is detected in 9, CN 1-0 is detected in 10 and CN 2-1 in 7 of the
galaxies and
all are new detections. We also report the first detection of HC$_3$N (10-9) emission
in Arp~220. The excitation of HNC and CN emission requires densities $n > 10^4$ cm$^{-3}$.
We compare their intensities to that of the usual high density tracer HCN.  
{\it The ${I({\rm HCN}) \over I({\rm HNC})}$ 1-0 and ${I({\rm HCN}) \over I({\rm CN})}$ 1-0
line intensity ratios vary
significantly, from 0.5 to $\gapprox 6$, among the galaxies.} This implies that the actual
properties of the dense gas is varying among galaxies who otherwise have similar
${I({\rm CO}) \over I({\rm HCN})}$ line intensity ratios.
We suggest that the HNC emission is not a reliable tracer of cold (10 K) gas at the center
of LIR galaxies, as it often
is in the disk of the Milky Way. Instead, the HNC abundance may remain substantial,
despite high gas temperatures, because the emission is emerging from regions
where the HCN and HNC formation and destruction processes are dominated by ion-neutral reactions
which are not strongly dependent on kinetic temperature.
We find five galaxies (Mrk~231, NGC~7469, NGC~7130, IC~694 and NGC~2623) 
where the ${I({\rm HCN}) \over I({\rm HNC})}$ intensity ratio is close to unity. 
Four are classified as active galaxies and one as a starburst. In other active galaxies, however,
the ${I({\rm HCN}) \over I({\rm HNC})}$ is $>4$.
The CN emission is on average a factor of two fainter than the HCN for the luminous IR galaxies,
but the variation is large and there seems to be a trend of reduced relative CN luminosity with
increasing
IR luminosity. This trend is discussed in terms of other PDR tracers such as the [C II] 158 $\mu$m
line emission. One object, NGC~3690, has a CN luminosity twice that of HCN and its ISM is thus
strongly affected by UV radiation. 
We discuss the ${I({\rm HCN}) \over I({\rm HNC})}$ and ${I({\rm HCN}) \over I({\rm CN})}$ line ratios
as indicators of starburst evolution. However, faint HNC emission is expected both in a shock
dominated ISM as well as for a
cloud ensemble dominated by dense warm gas in the very early stages of a starburst. Additional
information will help resolve the dichotomy.
}
\maketitle
\section{Introduction}

The polar molecule HCN (dipole moment 2.98 debye) is commonly used as a 
tracer of dense molecular gas, i.e. gas at $n({\rm H}_2) \geq 10^4$\,cm$^{-3}$. 
In particular in distant luminous (L$_{\rm IR}> 10^{11}$ L$_{\odot}$, LIRGs) and
ultraluminous (L$_{\rm IR}> 10^{12}$ L$_{\odot}$, ULIRGs) systems the HCN 1--0 line
is the prototypical tracer of 
dense gas content (e.g., Solomon \etal\ 1992; Helfer \& Blitz 1993; Curran \etal\ 2000 (CAB)). 
Solomon \etal\ (1992) find a tighter correlation between FIR and HCN luminosity than
the one found between FIR and CO. They suggest that, in general, the IR luminosities originate from
star formation rather than AGN activity in FIR luminous galaxies.  
The HCN to CO intensity
ratio, however, varies substantially (${1 \over 3}$ - ${1 \over 40}$) among luminous galaxies,
and it is unclear whether this difference can simply be interpreted as variations
in the dense gas content or is also due to abundance and/or excitation effects.
Apart from being collisionally excited, HCN may become excited via electron collisions
(at $X(e) \approx 10^{-5}$) or be pumped by 14\,$\mu$m continuum radiation through vibrational
transitions in its degenerate bending mode. It is also difficult to know if the gas
is really engaged in star formation, or if it is simply dense in response to being near
the central potential of the galaxy (e.g., Helfer \& Blitz 1993; Aalto \etal\ 1995)
where other mechanisms (AGN, turbulence etc) may heat the gas and dust.

In order to understand the activities in the centers of luminous galaxies it is essential
to also understand the prevailing conditions of the dense gas. Apart from observing higher
transitions of HCN it is important also to study the emission from other high density tracers.
One such tracer is the HNC molecule, the isomer of (and chemically linked to) HCN.
For example, at high temperatures HNC can be transferred into HCN via the reaction
${\rm HNC} + {\rm H} \to  {\rm HCN} + {\rm H}$. 
It is predicted, e.g.\ in (maybe oversimplified) chemical steady state models,
but also by shock models, that the ${{\rm HCN} \over {\rm HNC}}$ ratio increases with
increasing temperature 
and gas density (e.g., Schilke \etal\ 1992 (S92)). This is supported by the fact 
that the measured ${{\rm HCN} \over {\rm HNC}}$ abundance ratio is especially high in the vicinity of 
the hot core of Orion KL. Most of the temperature dependence is between 10 and 50 K,
after which there is a considerable flattening (S92).

Compared to these results, the ${{\rm HCN} \over {\rm HNC}}$ intensity ratios found (so far) in nearby
starburst galaxies are rather low (ranging from 1-5) closer to dark clouds than to
hot cores (H\"uttemeister \etal\ 1995 (H95)). This result is in apparent contradiction with
the idea that the gas is warm ($T \gapprox 50$ K) in the centers of starburst galaxies
(e.g., Wild \etal\ 1992; Wall \etal 1993). However, Aalto \etal\ (1995) suggest that
the dense cores of the
molecular clouds of the starburst NGC~1808 are cold (10 K) and thus these cores could
be responsible for the HNC emission in NGC~1808, but possibly also in other galaxies.

The radical CN is another tracer of
dense gas, with a somewhat lower (by a factor of 5) critical density than HCN. 
Observations of the CN emission towards the Orion~A molecular complex 
(Rodriguez-Franco \etal\ 1998) show that the morphology of the CN emission is dominated 
by the ionization fronts of the HII regions. The authors conclude that this molecule 
is an excellent tracer of regions affected by UV radiation. Thus, the emission from the
CN molecule should serve as a measure of the relative importance of gas in Photon Dominated
Regions (PDRs).

We have searched for HNC and CN emission in a sample of LIRG and ULIRG
galaxies with warm (${{f(60 \, \mu{\rm m})} \over {f(100 \, \mu{\rm m})}} \gapprox 0.75$)
FIR colours. 
We were interested to see whether the HNC emission would
be relatively fainter compared to the cooler, nearby objects studied by H95.
Is HNC a reliable cold gas tracer, or would we find evidence for the contrary?
We furthermore wanted to assess the relative importance of dense PDRs in these objects
through comparing the CN line brightness with that of HCN.
If indeed the HNC emission is a tracer of the amount of cold, dense gas, then perhaps an
anti-correlation between the CN and HNC emission is to be expected. Many of the galaxies
in the survey are powered by prodigious rates of star formation and thus a bright CN
line relative to that HCN is to be expected. Some of the galaxies are dominated by an AGN
where the expected CN brightness may also be high (e.g., Krolik \& Kallman 1983).

In Sect. 2, we present the observations and in Sect. 3 the results in terms of line
intensities and line ratios. In Sect. 4.1 we discuss the interpretation of the
HNC results and in Sect. 4.2 we discuss CN. In 4.3 possible connections to
starburst evolution and scenarios of the possibly dominating 
gas components are discussed.

\section{Observations}

We have used the SEST and OSO 20m to measure the HNC 1-0 (90.663 GHz) and the CN 1-0 113.491 GHz
(1-0, $J$=3/2-1/2, $F$=5/2-3/2) line intensity in
a selection of 13 LIRGs and ULIRGs. We also include observations of NGC~1808 which is of lower
luminosity.
The selected galaxies all have global ${I({\rm CO}) \over I({\rm HCN})}$ 1-0 intensity ratios
$\lapprox 15$ (apart from NGC~3256 and NGC~1808).
For the southern galaxies observed with SEST, we were also
able to measure the CN 2-1 line (226.874 GHz (2-1, $J$=5/2-3/2, $F$=7/2-5/2), 226.659 GHz
(2-1 J=3/2-1/2 F=5/2-3/2)). The CN 1-0 113.191 GHz line (1-0 J=1/2-1/2 F=3/2-3/2) is shifted
+806 \kms\ from the main line and we have obtained limits to its intensity in several cases.
For Arp~220 the two CN 1-0 spingroups are blended because the line is so broad. Thus, even in
the 1 GHz correlator backend
(see below) it was necessary to observe CN 1-0 at two different LO settings and then join the
spectra together to get enough baseline. 
For four galaxies the bandwidth was wide enough to also include the 90.983 GHz 10-9 line of HC$_3$N.

Observations were made in 1999 October (HNC, OSO), December (HNC, SEST) and 2000 June (CN, OSO),
August (CN, SEST). For OSO, the system temperatures were typically 300 K for HNC
and 500-600 K for CN. 
For SEST, typical system temperatures were 230\,K for the HNC 
measurements and 400\,K for both the 113\,GHz and the 226\,GHz CN
observations.
Pointing was checked regularly on 
SiO masers and the rms was found to be 2$''$ for OSO, and 3$''$ for SEST.
Arp~220 was observed both with OSO (CN) and SEST (HNC). We have also measured the 115 GHz CO 1-0
and the HCN 1-0 lines for some galaxies where we did not have values from the literature.
Beamsizes and efficiencies are shown in Table~1.
For the OSO observations a 500 MHz filterbank was used for backends for all observations, and
for some a 1 GHz autocorrelator was also used. For the SEST observations we alternated between
a 500 MHz and 1 GHz backend depending on whether simultaneous observations with the 1 and
3 mm receiver were taking place. We used the software package xs (written by P. Bergman) to
subtract baselines and add spectra.

\begin{table}
\caption{\label{beam} Beamsizes and efficiencies}
\begin{tabular}{lccccc}
Transition & $\nu$ [GHz] & \multicolumn{2}{c}{HPBW [$''$]} & \multicolumn{2}{c}{$\eta_{\rm mb}$}\\
 & & OSO & SEST & OSO & SEST\\
 \hline \\
HCN 1-0 & 88.6 & 44 & 57 & 0.59 & 0.75\\
HNC 1-0 & 90.6 & 42 & 55 & 0.59 & 0.75\\
CN 1-0 & 113.5 & 34 & 46 & 0.50 & 0.70\\
CO 1-0 & 115.2 & 33 & 45 & 0.50 & 0.70\\
\hline \\
\end{tabular} \\
\end{table}

\section{Results}
\subsection{HNC line intensities and ratios}

\begin{table*}
\caption{\label{res} Integrated Line Intensities$^{\rm a}$}
\begin{tabular}{lcccccccc}
Galaxy & Telescope & $v_c$ & $I$(HNC) 1-0$^{\rm b}$ & $I$(CN) 1-0$^{\rm c}$ & $I$(CN)2-1$^{\rm d}$
& 5/2 - 3/2 & 3/2 - 1/2 & $I$(CO) 1-0  \\
\hline \\
& & \kms\ & K \kms\ & K \kms\ & K \kms\ & K \kms\ & K \kms\ & K \kms \\
\hline \\

Arp~220 & OSO & 5400 & $\dots$ & $2.0 \pm 0.3$ & $\dots$ & $\dots$ & $\dots$ & $15.0 \pm 2.0^{\rm g}$ \\
  "     & SEST & 5400 & $0.95 \pm 0.2$ & $\dots$ & $0.65 \pm 0.2^{\rm h}$ & $\dots$ & $\dots$ & $13.0 \pm 2.0$ \\
IC~694$^{\rm f}$ & OSO & 3100 & $0.75 \pm 0.2$ & $\lapprox 0.85$ & $\dots$ & $\dots$ & $\dots$ & $14.0 \pm 1.0$\\
NGC~3690$^{\rm f}$ & OSO & 3050 & $\dots$ & $1.5 \pm 0.3$ & $\dots$ & $\dots$ & $\dots$ & $11.0 \pm 1.0$\\
Mrk~231 & OSO & 12650 & $0.5 \pm 0.1$ & $0.5 \pm 0.2$: & $\dots$ & $\dots$ & $\dots$ & g\\
Mrk~273 & OSO & 11320 & $\lapprox 0.25$ & $\lapprox 0.25$ & $\dots$ & 
$\dots$ & $\dots$ & g\\
NGC~34 & SEST & 5930 & $\lapprox 0.3$ & $\lapprox 0.3$ & $0.25 \pm 0.05$ & $\dots$ & $\dots$ & $5.5 \pm 0.5^{\rm g}$\\ 
NGC~1614 & SEST & 4500 & $\dots$ & $0.6 \pm 0.3$: & $0.5 \pm 0.1$  & $\dots$ & $\dots$ & $9.0 \pm 0.5 $\\
NGC~2146 & OSO  & 900 & $\dots$ & $1.2 \pm 0.3$ & $\dots$ & $\dots$ & $\dots$ & $43.0 \pm 1.0$ \\
NGC~2623 & OSO & 5500 & $0.6 \pm 0.15$ & $\lapprox 0.6$ & $\dots$ & $\dots$ & $\dots$ & $13.0 \pm 0.5$ \\
NGC~6240 & SEST & 7335 & $0.4 \pm 0.2$: & $0.7 \pm 0.1$ & $1.3 \pm 0.1$ & $0.9 \pm 0.1^{\rm e}$ &
 $0.4 \pm 0.1^{\rm e}$ & $15.5 \pm 0.5$\\
NGC~3256 & SEST & 2800 & $0.6 \pm 0.05$ & $1.2 \pm 0.05$ & $0.55 \pm 0.05$ & $0.4 \pm 0.05$ & 
$0.15 \pm 0.05$  & $45 \pm 1.5$ \\
NGC~7130 & SEST & 4840 & $0.4 \pm 0.05$ & $0.5 \pm 0.1$ & $0.45 \pm 0.05$ & $0.3 \pm 0.05$ &
$0.15 \pm 0.05$ & $9.5 \pm 0.5^{\rm g}$ \\
NGC~7469 & OSO & 4960 & $0.7 \pm 0.1$ & $0.8 \pm 0.1$ & $\dots$ & $\dots$ & $\dots$ & g \\
\hline \\
NGC~1808 & SEST & 960 & $1.2 \pm 0.1$ & $3.8 \pm 0.1$ & $2.0 \pm 0.1$  & $1.4 \pm 0.1$  &
$0.7 \pm 0.1$ & $66 \pm 0.5$  \\
\hline \\
\end{tabular} \\
a): Integrated line intensities are in $T_{\rm A}^*$ and in K \kms, and upper limits are all $3\sigma$.
SEST, $T_{\rm mb}$: 27 Jy K$^{-1}$ (115 GHz), 41 Jy K$^{-1}$ (230 GHz) for point sources;
OSO, $T_{\rm mb}$: 27 Jy K$^{-1}$ (115 GHz) for point sources. \\
b): In Arp~220 we also detect the 90.98 GHz (10-9) line of HC$_3$N with the intensity
$0.4 \pm 0.15$.  For NGC~3256
the ($3 \sigma$) limit to the line is 0.18, for NGC~7130 it is 0.15,
and for NGC~1808 we have
a very tentative detection with the intensity $0.2 \pm 0.1$.
For Arp~220 we also find an integrated HCN intensity in the SEST beam of $2.5 \pm 0.4$. \\
c): This is only the integrated intensity of the 1-0 $J$=3/2-1/2 line. To get the total
1-0 intensity multiply with 1.3 (in the optically thin case). For Arp~220 both lines
are blended and included in the integrated line intensity listed here.\\
d): The total integrated intensity from the two 5/2 - 3/2 and 3/2 - 1/2 spingroups together.
The following two columns contain the integrated intensity in each spin group whenever it
was possible to measure. \\
e): Line blending is severe therefore the intensity ratio was estimated from fitting
of two gaussians with fixed linewidths of $\Delta V = 450$ \kms\ at $v_1=7200$ \kms and
$v_2=7500$ \kms.\\
f): IC~694 and NGC~3690 are also known as the merger Arp~299.\\
g): See CAB. \\
h): very broad lines, may be affected by unknown baseline errors.

\end{table*}

\begin{table*}
\caption{\label{res} Line Ratios and FIR luminosities}
\begin{tabular}{lllccccccc}
Galaxy & type$^{\rm a}$ & ${{\rm CO} \over {\rm CN}}$ 1-0 &  ${{\rm CO} \over {\rm HCN}}^{\rm b}$ 
& ${{\rm HCN} \over {\rm CN}}$ 
& ${{\rm HCN} \over {\rm HNC}}$ & ${{\rm HNC} \over {\rm CN}}$ & CN ${5/2 - 1/2 \over 3/2 - 1/2}$ & 
CN ${2-1\over 1-0}^{\rm c}$ & log L(FIR)$^{\rm d}$
  \\
\hline \\
Arp~220 & HII & 7.5 & 8 & 1 & 1.6 & 0.5 & $\dots$ & 0.15 & 12.13 \\
IC~694 & HII & $\gapprox 17$ & 11 &$\gapprox 1.2$ & 1.0$^{\rm e}$ & $\gapprox 1$ & $\dots$ & $\dots$ & 11.77 \\
NGC~3690 & HII & 6 & 13 & 0.5 & $\dots$ & $\dots$ & $\dots$ & $\dots$ & $\dots$ \\
Mrk~231 & Sey 1 & 8 & 5 & 1.6 & 1 & 1.6 & $\dots$ & $\dots$ & 12.37\\
Mrk~273 & Sey 2 & $\gapprox 10$ & 2$^{\rm g}$ & $\gapprox 5$ & $\gapprox 4$ & $\dots$ & $\dots$ & $\dots$ & 12.04\\
NGC~34 & Sey 2/HII & $\gapprox 17$ & 4 & $\gapprox $4 & $\gapprox $4 & $\dots$ & $\dots$ & 
$\gapprox 0.35$ & 11.16 \\
NGC~1614 & HII & 12 & 14 & 1 & $\dots$ & $\dots$ & $\dots$ & 0.35 & 11.43\\
NGC~2146 & HII & 28 & 16 & 1.8 & $\dots$ & $\dots$ & $\dots$ & $\dots$ & 11.00\\
NGC~2623 & LINER & $\gapprox 22$ & 12  & $\gapprox 2$ & 1.4 & $\gapprox 1.3$ & $\dots$ & $\dots$ & 11.49\\
NGC~6240 & Seyfert 2 & 18 & 9 & 2 & 2.7 & 0.7 & 2.1 & 0.6 & 11.69\\
NGC~3256 & HII & 29 & 16 & 2 & 3 & 0.7 & 2.6 & 0.14 & 11.52\\
NGC~7130 & Sey 2/HII & 14 & 13 & 1.1 & 1.2 & 1 & 2.4 & 0.25 & 11.26\\
NGC~7469 & Sey 1/HII  & 11 & 8$^{\rm h}$ & 1.4 & 1.2 & 1 & $\dots$ & $\dots$ & 11.41\\
\hline \\
NGC~1808$^{\rm f}$ & HII & 13 & 18 & 0.7 & 2 & 0.4 & 2.2 & 0.25 & 10.48\\
\hline \\
\end{tabular}\\
a): Type of central activity that is suggested to dominate the IR luminosity.\\
b): The HCN data come from CAB apart from for: IC~694 \& NGC~3690 (Arp~299)
(Solomon \etal\ 1992);
NGC~1614 (Aalto \etal\ 1995); NGC~2623, NGC~6240 (Bryant 1996); NGC~3256 (Aalto \etal\ 1995).
The Bryant (1996) ratios are measured with the OVRO interferometer, but since both the CO
and HCN emission is compact for both NGC~2623 and NGC~6240 the line ratios are global and can be
compared with single dish ratios.
For Arp~299 (IC~694 \& NGC~3690) the situation is more complicated since the high density tracer emission is emerging
from compact structures, while a fraction of the CO emission comes from more extended gas
(e.g., Aalto \etal\ 1997). In (Solomon \etal\ 1992) HCN was measured in a 28$''$
beam (IRAM).\\
c): We have not mapped the CN line emission which is of course necessary for an accurate
determination of the CN ${2-1 \over 1-0}$ line ratio. For most galaxies we expect the CN distribution
to be effectively a point source in the beam --- except for NGC~1808 where we adopt a 
source size of 18$''$ (Aalto \etal\ 1995).\\
d): The FIR luminosities come from Bryant (1996) (apart from  NGC~1808 (Aalto \etal\ 1994);
NGC~34 (CAB); NGC~3256, NGC~7130, NGC~2146 (Aalto \etal\ 1991))\\
e): We have assumed that the HNC emission is a point source (see footnote b) and recalculated
the line intensity to that of the 28$''$ IRAM HCN beam.  \\
f): Line ratios are not global, but apply only for the inner 2-3 KPC.\\
g): We use the temperature line ratio of 2 instead of the integrated line ratio of 1 given in
CAB (because the HCN line appear too broad). \\
h): There is an error in table 4 of CAB. The correct ratio (8) can be found in table 3 in CAB.

\end{table*}

The line intensities are presented in Table~2 and the ratios in Table~3. CO spectra are presented
in Figure 1 and HNC spectra (plus a SEST HCN spectrum for Arp~220) in Figure 2. For some of the
galaxies the
center velocity $v_c$ is put to 0. This is because the observer in that case chose to set the velcocity
to zero and work with the redshifted frequency instead.  The velocity used to calculate the redshifted
frequency (thus the velocity the band was centered on) can be found in Table~2. 
All galaxies, except NGC~1808, have FIR luminosities $L_{\rm FIR} \gapprox 10^{11}$ L$_{\odot}$
(see Table~3).
Five of the investigated sources (Mrk~231, NGC~7469, NGC~2623, IC~694 and NGC~7130) have global
${{\rm HCN} \over {\rm HNC}}$
luminosity ratios close to unity. The rest have ratios ranging from 2 to $\gapprox 6$. The HNC luminous
objects are all AGNs (three Seyferts, one LINER (NGC~2623)) except for IC~694 which we suspect is
a starburst (e.g., Polatidis \& Aalto 2000), but there are also Seyfert galaxies
with faint HNC emission (such as Mrk~273, NGC~6240 and NGC~34). 
We do not find that an increase in FIR luminosity is followed by an increase in
${{\rm HCN} \over {\rm HNC}}$
line intensity ratio. Instead, there might be a weak trend to the opposite and two (Mrk~231 and Arp~220)
of the three ULIRGs in our sample have relatively bright HNC emission (see Sect. 3.1.1).

\subsubsection{Comparisons with other HNC surveys}

In the H95 HNC survey of nearby starburst galaxies, the
majority of the sources show ${{\rm HCN} \over {\rm HNC}}$ line ratios greater than or equal to two.
Three objects have $I$(HCN)$ \approx I$(HNC): the nearby starburst NGC~253, the nearby post-starburst
NGC~7331 and the Seyfert NGC~3079.
For most of the galaxies in that sample, the line ratios are not global, but
reflect the conditions in the inner  0.2 - 1 kpc of the galaxy. Since both the HCN
and CO emission for most of our sample galaxies comes from the inner kpc
(e.g., Bryant 1996; Scoville \etal\ 1997; Downes \& Solomon 1998; Bryant \& Scoville 1999), it
is meaningful to compare the ${{\rm HCN} \over {\rm HNC}}$ ratios of the two samples. 
In H95 the average ${{\rm HCN} \over {\rm HNC}}$ line intensity ratio is 2 for 14 galaxies
(excluding limits and their value for Arp~220 (see below)). 
In our sample, the ratio is somewhat
lower, 1.6, when only detections are included (but when limits are included the ratio 
increases to 2). Joining the H95 galaxies with ours in one sample we still can find
no strong trend in the line ratio with increasing FIR luminosity. We note, however,
that we have a larger number of galaxies with ratios close to unity (6) compared to
H95 (3) despite our smaller sample.

We observed HNC in Arp~220 to see if we could reproduce the remarkable result in H95 that
HNC was brighter than HCN. We observed HNC on two consecutive days, with observations
of HCN in between, but could not confirm the bright HNC emission found earlier.
Instead we obtain an ${{\rm HCN} \over {\rm HNC}}$ intensity ratio of 1.4, which is more
consistent with results found in other galaxies. 

\subsubsection{HC$_3$N}

For some galaxies the bandwidth was large enough to include the HC$_3$N 10-9 line, shifted in velocity
by -1000 \kms\ from the HNC line.
The line is detected in Arp~220 at 40\% of the HNC 1-0 intensity. In NGC~1808 the line is tentatively
detected at 16\% of the HNC 1-0 intensity and in NGC~7130 and NGC~3256 we have upper limits to the 
line (see footnote to Table~2). 

\subsection{CN line ratios and intensities}

{\it CN 1-0:} The line intensities are presented in Table~2 and the ratios in Table~3. The CN 1-0
spectra are displayed in Figure 3. For most of
the galaxies we only detect one of the spingroups in the 1-0 transition.
Because of the broad line in Arp~220 the two groups are however blended --- even though they are
separated by 800 \kms. Also for NGC~6240 the second spingroup is somewhat blended with the first one.

The ${{\rm HCN} \over {\rm CN}}$ 1-0 integrated line ratio varies substantially: from 0.5 to
greater than four.
$I$(CN)$ \gapprox I$(HCN) in three
objects (Arp~220, NGC~3690 and NGC~1808), while CN remains undetected in IC~694, Mrk~273, NGC~34 and
NGC~2623.
We also find three galaxies (Mrk~231, NGC~2623, and IC~694) where
$I$(CN) $\lapprox I$(HNC). 
In four galaxies (Arp~220, NGC~1808, NGC~3256, NGC~6240) $I$(HNC) $<$ $I$(CN)
--- three are starburst galaxies and NGC~6240 is classified as a Seyfert. In H95, CN was brighter
than HNC in all four sample galaxies where CN was measured.

The ${{\rm HCN} \over {\rm CN}}$ line intensity ratio seems to increase slightly with FIR luminosity.
Dividing our galaxies into two luminosity bins we find that the average ratio of 1.5 for the lower
luminosity galaxies and 2.3 for the higher luminosity objects.
This, somewhat surprising, result is discussed in Sect. 4.2.
We included NGC~3690 in the lower luminosity batch since most of the FIR emission is
believed to originate in IC~694.

{\it CN 2-1:} The 2-1 spectra are presented in Figure 4. Both 2-1 spingroups are
detected in four of the cases (in NGC~6240 the blending is severe).
In NGC~34 and in NGC~1614 we detect only the brighter $J$=5/2-3/2 transition.
The CN 2--1 spectrum of Arp~220 shows a much weaker line than
the 1--0 spectrum, and emission is only detected in the lower
velocity part ($v_c \approx 5200$ \kms) of the spectrum.
Also the CN 1-0 line
peaks around 5200 \kms. The CO emission peaks at $v_c \approx 5400$ \kms. This implies that
most of the CN emission emerges from the western nucleus (see Scoville \etal\ 1997).
The HC$_3$N line on the other hand appears to peak at $v_c =5550$ \kms, and thus most of the emission
likely emerges from the eastern nucleus. This tentative velocity difference should be investigated at
higher resolution.

\subsubsection{The CN ${2-1 \over 1-0}$ ratio}

The CN ${2-1 \over 1-0}$ line intensity ratio (see footnote to Table~3 on how it was calculated)
suggests the CN emission being subthermally excited apart from in NGC~6240 (and NGC~34
with an upper limit to the CN 1-0 emission). In general the ratio is lower than 0.4
indicating gas densities $\lapprox 5 \times 10^4$ $\cmmd$ (Fuente \etal\ 1995) which are about
one order of magnitude below the critical density.

\subsubsection{The CN spingroup ratio}

We can use the 1mm spingroup ratio to estimate the average optical depth
of the two lines. The ratio between the lines is quite large, $>$2 for all cases
where it can be measured. The spingroup ratios are very accurate since they were
obtained at the same time (apart from possible baseline errors). In NGC~6240 the line
blending is so severe that the ratio
is difficult to determine.  The LTE ratio
should be close to 1.8 (Bachiller \etal\ 1997) and ratios around 2 thus indicate
that the lines are of low optical depth.
For most galaxies, the other spingroup of the 1-0 line (around 113.191 GHz ($J$=1/2-1/2)) was outside
the observed bandwidth. In the optically thin case, this line is about 1/3 of the
$J$=3/2-1/2 at 113.490. For Arp~220, the two lines are blended (see above) and we fitted two gaussians
each centered on one of the spingroups. The fits shows that a line intensity ratio between the
two lines of 1/3 is possible.

\begin{figure*}
\resizebox{16cm}{!}{\includegraphics{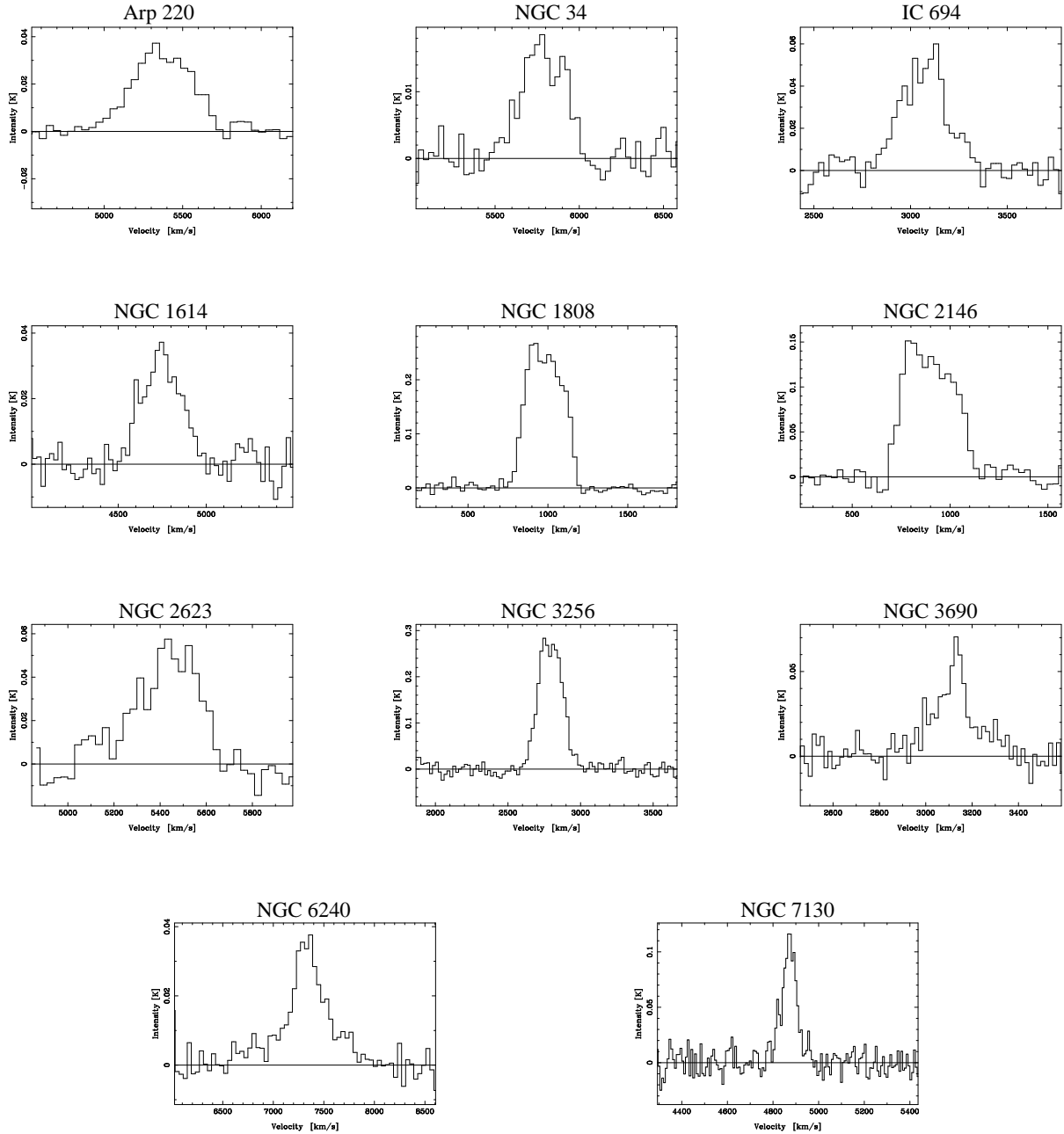}}
\caption{\label{something} CO spectra of all galaxies, apart from MRK~231, MRK~273 and NGC~7469
which can be found in CAB. The scale is in
$T_{\rm A}^*$. The velocity resolution ranges from 10 TO 50 \kms. It was selected to be 10\% of
the FWHM of the line itself.}
\end{figure*}

\begin{figure*}
\resizebox{16cm}{!}{\includegraphics{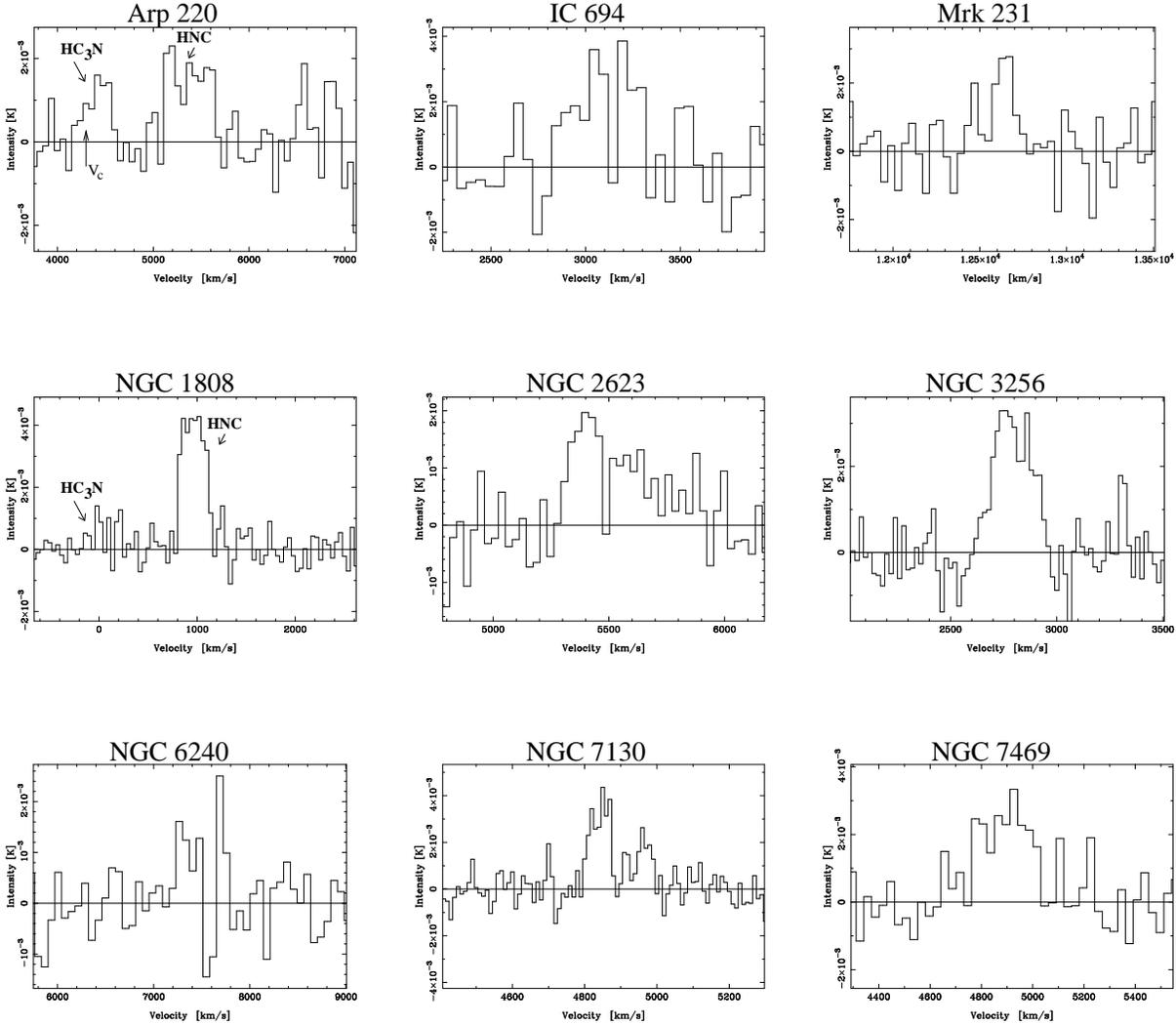}}
\caption{\label{something}  HNC 1-0 spectra for the galaxies for which we claim a 
detection. The scale is in $T_{\rm A}^*$. The 10-9 HC$_3$N line is indicated in the
spectra for Arp~220 and NGC~1808. The velocity resolution ranges from 10 to 50 \kms. It was selected
to be 10\% of the FWHM of the line itself, but for a few cases the S/N of the spectrum required
further smoothing. }
\end{figure*}

\begin{figure*}
\resizebox{16cm}{!}{\includegraphics{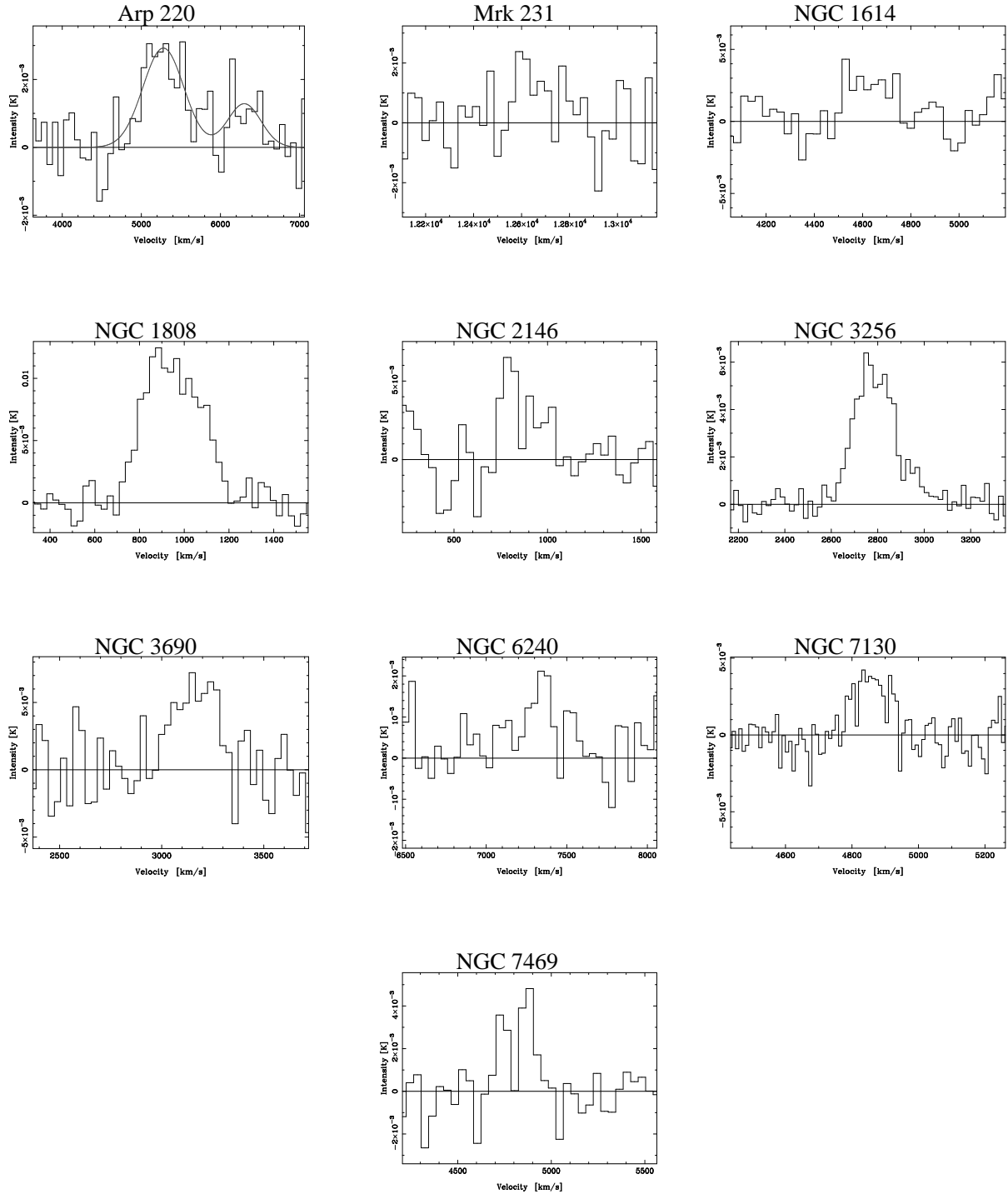}}
\caption{\label{something}  CN 1-0 spectra for the galaxies with detections.
The scale is in $T_{\rm A}^*$. The spectrum for Arp~220 shows the Gaussian fit for both 
spingroups, since the line is broad enough for them to be blended. The velocity resolution ranges from 10 to 50 \kms. It was selected
to be 10\% of the FWHM of the line itself, but for a few cases the S/N of the spectrum required
further smoothing.}
\end{figure*}

\begin{figure*}
\resizebox{16cm}{!}{\includegraphics{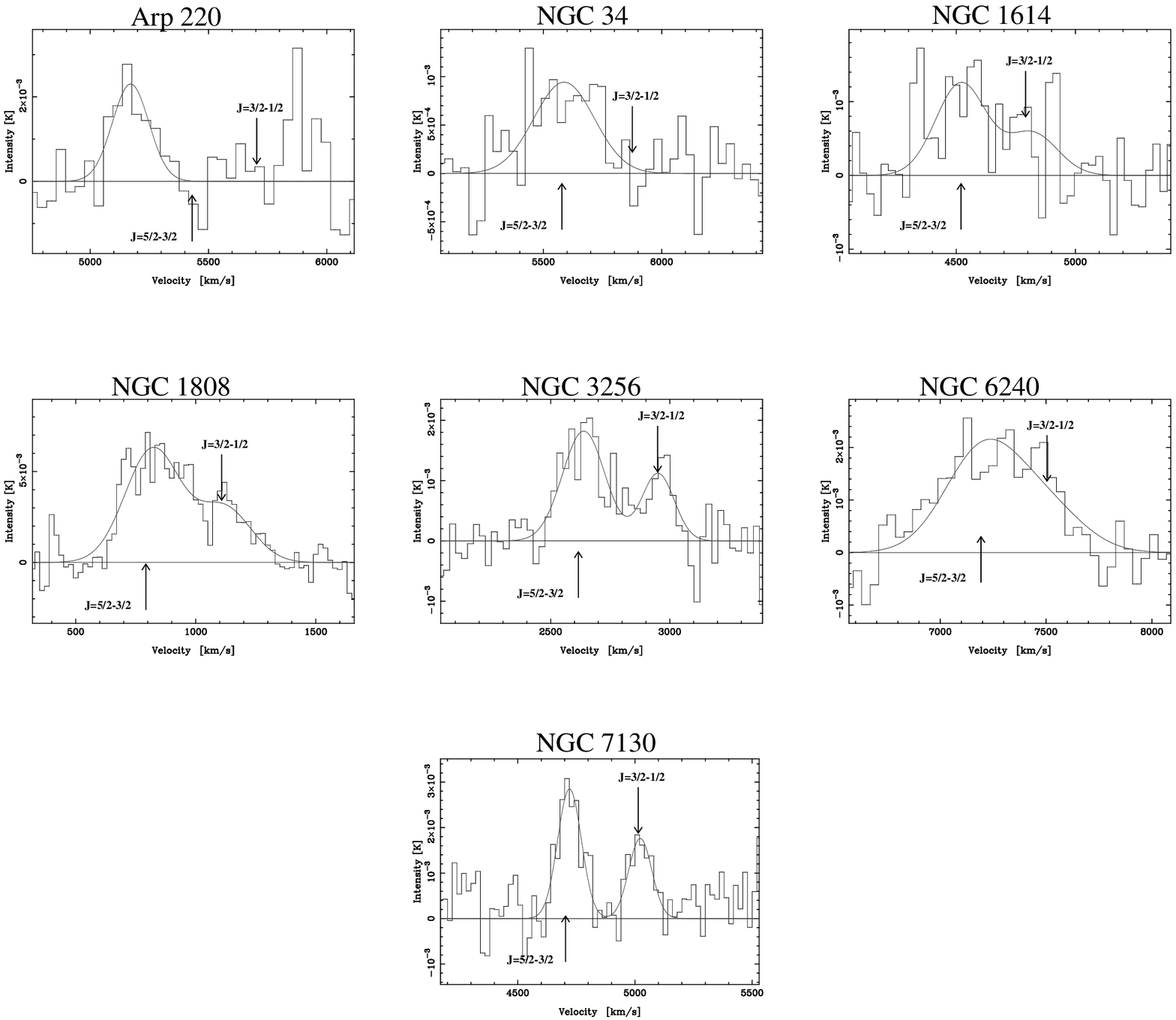}}
\caption{\label{something}  CN 2-1 spectra for the galaxies with detections.
The scale is in $T_{\rm A}^*$. The band is centered in between the two spingroups,
apart from Arp~220, where the spectrometer was centered on the $J$=5/2-3/2 line.
The two spingroups are marked with arrows and are blended for all galaxies, apart from
NGC~7130. FOR Arp~220 emission is not detected at the line center (marked by the arrow)
instead, there is a tentative detection at $V$=5170 \kms\ (blueshifted from $V_c$ by
approximately 250 \kms). The velocity resolution ranges from 10 to 50 \kms. It was selected
to be 10\% of the FWHM of the line itself, but for a few cases the S/N of the spectrum required
further smoothing.}
\end{figure*}

\section{Discussion}

We initially expected that the relative HNC luminosity would be lower in 
our sample of warm, luminous galaxies compared to the H95 sample. Instead
(see Sect. 3.1.1) the very luminous galaxies even seemed to be somewhat more
luminous, on average, in HNC even though this change is
not statistically significant. A possibility could be that the telescope beam is
picking up more extended, cooler gas in the distant, more luminous, galaxies. However, 
most of them are known to have very compact molecular cloud distributions and it is
unlikely that significant HNC emission emerges from the outskirts of the galaxies.
{\it Our results seem to challenge the notion of HNC as a reliable tracer of cold gas.}

Furthermore, the variation in ${{\rm HCN} \over {\rm HNC}}$ line ratio is large among the
galaxies that otherwise have similar properties.
For example, despite having similar ${{\rm CO} \over {\rm HCN}}$ 1-0 intensity ratios,
Mrk~273 and Mrk~231 have quite
different ${{\rm HCN} \over {\rm HNC}}$ and ${{\rm HCN} \over {\rm CN}}$ line intensity ratios.
The ${{\rm HCN} \over {\rm HNC}}$ ratio of Mrk~231
is close to unity, while HNC is not detected in Mrk~273 resulting in a
${{\rm HCN} \over {\rm HNC}}$ ratio
$\gapprox 6$. It is unlikely that the dense gas is cold in Mrk~231 but warm in Mrk~273
(both are hot, ultraluminous AGN/Starburst mergers) and therefore the interpretation
of HNC needs to be reevaluated.

\subsection{Why is HNC often bright in centers of galaxies?}

There are a number of possible explanations. Let us examine them one by one and see
how they can be tested:

1. {\it The dense gas is cold}. 

Is it possible that a significant fraction of the
dense gas {\em is} cold in centers of starburst galaxies? Possibly. Aalto \etal\ (1994) discuss the
presence of cold dense gas in the ``mild'' starburst galaxy NGC~1808. The high dust
temperatures observed towards NGC~1808 could be explained by clouds with hot surfaces and
cold interiors. In this scenario, the HNC 1-0 emission would emerge from the cold cloud cores, 
while HCN 1-0 emission also would come from the outer, warmer parts.
Gas at densities $>10^5$ $\cmmd$ should become thermalized with the dust.
Thus, if the HNC emission is coming from dense, cold (10-20 K) gas there should be
submm dust continuum emission associated with it. Therefore, a study of the submm and mm excess
in conjunction with the strength of the HNC emission  should be quite interesting.
However, the most HNC luminous object in our sample, Mrk~213, shows very weak mm thermal dust 
emission and Braine \& Dumke (1998) find a dust temperature of 70 K --- the kinetic temperature
of the associated dense gas should be at least as high. This is not consistent with the idea that
the HNC emission arises from a cold component.
Based on ISOPHOT data Klaas \etal\ (1997) find that the bulk of the dust mass in Arp~220 is at a
temperature of about 50 K. Thus, the bright HNC 1-0 emission is unlikely originate in cold cloud
cores.
 
We conclude that for objects like Mrk~231 and Arp~220, where there is no mm excess emission,
and where the overall dust temperature is high, HNC 1-0 does not trace cold dense gas. Instead
one or several of the scenarios below must apply. However, for less extreme objects such as
NGC~1808 there could be enough mass in cold cores that at least a fraction of the
HNC emission could emerge from them.

2. {\it Chemistry}. 

Steady state chemistry models by S92 show both
a temperature and a density dependence in the ${[{\rm HCN}] \over [{\rm HNC}]}$ abundance ratio.
For example, there is a very
significant difference between $n$(H$_2$)=$10^4$ $\cmmd$ and $10^7$ $\cmmd$.  For $T_{\rm kin}$= 50 K, 
${[{\rm HCN}] \over [{\rm HNC}]}$=0.8
for $10^4$ and ${[{\rm HCN}] \over [{\rm HNC}]}$=67 for $10^7$. If the bulk of the HCN and HNC
emission is emerging from gas
of densities $10^4 - 10^5$ then the relative HNC abundance there may be substantial, despite the
high temperature. The reason for this is that at lower densities reactions with HCNH$^+$ (HCN and
HNC reacts with H$_3 ^+$ to form HCNH$^+$) become
more important. The ion abundance is higher and once HCN and HNC become protonised, HCNH$^+$
will recombine to produce either HCN or HNC with 50\% probability. At higher densities, the
ion abundance is likely lower and reactions like HNC + O $\rightarrow$ CO + NH become more important
at high temperatures. This scenario is interesting since the electron and ion abundance is likely higher 
in PDRs. Therefore, in a PDR chemistry, the connection between HNC abundance and kinetic temperature
may also be weak since we there expect the HCNH$^+$ reactions to be important.
Since the CN ${2-1 \over 1-0}$ ratios we measure indicate subthermal excitation it is reasonable
to assume that most of the HCN,HNC and CN emission is indeed emerging from gas where the density
is below $10^5$ $\cmmd$.

3. {\it Optical depth effects}

In Orion, S92 find peak-to-peak {\it intensity} ratios between ${{\rm HCN} \over {\rm HNC}}$ of 
3 to 4
towards the hot core and ridge. However, the abundance ratio is much higher, $\approx$80.
Thus, it is possible that the fairly bright HNC emission in some galaxy centers is caused by
optical depth effects. In objects where the HCN emission is subthermally excited (like in
Mrk~231) the optical depth of the 1-0 line could be quite high and perhaps explain part of the
apparently too-bright HNC.

4. {\it IR pumping}

Both HCN and HNC may be pumped by an intense mid-IR radiation field boosting
the emission from low density regions.
There has been no direct evidence IR pumping is dominating the HCN excitation
in external galaxies.
However, Barvainis \etal\ (1997) suggest IR pumping as a possible mechanism behind
the HCN emission
of the Cloverleaf quasar. Ultraluminous galaxies, such as Mrk~231 and Arp~220, have
central mid-IR sources with optically thick radiation temperatures well in excess of
those necessary to pump the HCN molecule (Soifer \etal\ 1999).
For HNC the coupling to the field is even
stronger than for HCN, thus increasing the probability for IR pumping in extreme
galaxies, such as Mrk~231. Even if the HNC abundance is lower compared to HCN the
HNC emission may have a higher filling factor due to the IR pumping (if it allows
emission from gas clouds otherwise at too low density to excite the HNC molecule).
A comparative excitation study of HCN and HNC would help cast light on this issue.
Of course, the HNC abundance must be high enough so that the pumping can be effective
and it must thus happen in regions where the chemistry is dominated by ion-neutral
reactions.

\subsection{Interpreting the CN emission}

\subsubsection{CN chemistry}

Studies of Galactic molecular clouds have shown that the ${[{\rm CN}] \over [{\rm HCN}]}$ abundance
ratio is increasing in the outer regions of UV irradiated clouds (e.g., Greaves \& Church 1996;
Rodriguez-Franco \etal\ 1998). The abundance of the CN radical becomes enhanced
at the inner edge of a PDR
(at an $A_{\rm V}$ of about 2 magnitudes) via the reaction ${\rm N} + {\rm C}_2 \to 
{\rm CN} + {\rm C}$ or via ${\rm N} + {\rm CH} \to {\rm CN} + {\rm H}$. At larger
depths into the cloud the CN abundance radically declines
and the ${[{\rm HCN}] \over [{\rm CN}]}$ abundance ratio increases (Jansen \etal\ 1995). Most of
the CN is
present in a part of the cloud where the abundance of free electrons is rather large,
$X(e) \approx 3 \times 10^{-5}$. CN is also a photodissociation product of HCN.
Thus {\it the CN abundance should be favoured in a molecular cloud ensemble dominated by
PDRs}. 
Furthermore, chemical models (e.g., Krolik \& Kallman 1983; Lepp \& Dalgarno 1996) show that the CN 
abundance should also be enhanced when the X-ray ionization rates are high --- as might
be the case near an AGN.

It is of course difficult to translate a measured ${{\rm HCN} \over {\rm CN}}$ 1-0 line
intensity ratio to an
abundance ratio between the two species. 
The spingroup line ratios (see Sect. 3.2.1.) show that the CN 2-1 emission is optically
thin for most galaxies we have measured. The emission of the CN molecule is distributed
in a greater number
of transitions than HCN, thus the optical depth per transition is often lower for CN,
reducing the intensity per line. The CN luminosity then becomes a measure of 
the total number of CN molecules (at least in a comparative sense, given a constant
excitation situation from galaxy to galaxy). For the CN 1-0 line we have only information
for Arp~220 where the relative faintness of the second spingroup suggests that the optical
depth of the 1-0 line is also low. So, the measured (total) ${{\rm HCN} \over {\rm CN}}$
line intensity ratio
will give a reasonable idea of the abundance ratio if also the HCN line is close to being
optically thin (and a lower limit to the abundance ratio if it is not) and if the same
excitation temperature can be assumed. The critical density
of the CN line is lower by a factor of a few, so its $T_{\rm ex}$ is likely somewhat higher.

\subsubsection{AGNs or starbursts?}
There are only a few galaxies where $I$(CN) $\gapprox I$(HCN) (Arp~220, NGC~3690 and NGC~1808) ---
all three are starburst galaxies. For these galaxies $N$(CN) is greater than $N$(HCN) and
the dense gas is to a large degree affected by photodissociation.
The most extreme example is NGC~3690 where the CN line is on average more than a factor of two
brighter than HCN. In Arp~220 the CN appears to be bright only towards the western of the
two nuclei (Sect. 3.2). This is interesting, since the situation is similar for the
IC~694/NGC~3690 system: bright CN emission towards one of the galaxy nuclei only. It is
interesting to speculate on whether the burst towards NGC~3690 is older than the one in
IC~694 because of the development of the PDRs. However, the IC~694 burst is more compact
and it is possible that the actual properties of the ISM are intrinsically different. 

We were surprised to find that CN was difficult to detect in several of the brightest
galaxies like Mrk~273, NGC~2623, NGC~6240, IC~694 and NGC~34. Four of these galaxies
are AGNs and it is tempting to speculate that the CN deficiency is related to the nuclear
activity. This seems contrary to models (see above) which predict an increase in the CN
abundance in an X-ray chemistry. High resolution studies of nearby systems which contain
both starburst and AGN activity (like NGC~1068) will reveal whether CN emission is associated
with one or both of the activities.

\subsubsection{The CN and the [C II] 158 $\mu$m line}

In ULIRGs such as Arp~220 and Mrk~231 the [C II] 158 $\mu$m fine structure line is
found to be abnormally faint compared to other, less FIR luminous, starburst galaxies
like NGC~3690 (e.g., Luhman \etal\ 1998). This is interesting, since one would expect the
emission from a standard PDR tracer, like the [C II] line, to be bright in a galaxy that
is believed to be powered largely by mighty starbursts. 
Malhotra reports a decreasing trend in ${{F_{\rm [C II]}} \over {F_{\rm FIR}}}$ with increasing
${f(60) \over f(100)}$ $\mu$m flux ratio.

Several possible explanations for the [C II] faintness are brought forward (e.g., Malhotra
\etal\ 1997; Luhman \etal\ 1998; van der Werf 2001). The PDRs may be
quenched in the high pressure, high density environment in the deep potentials of the
ULIRGs and the HII regions exist in forms of small-volume, ultracompact HII regions
that are dust-bounded.
The [C II] line may become saturated either in low density ($n \propto 10^2$ $\cmmd$)
regions of very high UV fields ($G_0 \propto 10^3$) or in dense ($n \propto 10^5$ $\cmmd$)
regions of more moderate UV fields ($G_0 = 5-10$). A soft UV field from an aging starburst
is another possibility. A higher dust-to-gas ratio would also decrease the expected
${F_{\rm [C II]} \over F_{\rm FIR}}$ ratio.

The molecular ISM of ULIRGs seems to be characterized by subthermally excited CO and
very bright emission from HCN (e.g., Downes \& Solomon 1998; Aalto \etal\ 1995; Solomon \etal\ 1992). 
Crudely, this can be modelled as dense clouds ($n = 10^4 - 10^5$ $\cmmd$)
embedded in a low density ($n= 100 - 500$ $\cmmd$) continous medium. This simple scenario
may fit well with the two scenarios resulting in saturated [C II] emission. 

We have three ULIRGs in our sample: Arp~220, Mrk~231 and Mrk~273  their ${{\rm HCN} \over {\rm CN}}$ 
line intensity ratio changes from 1 to $\gapprox 6$. The deficiency of CN in Mrk~273 is consistent
with the lack of [C II] emission and can be an indication that the PDRs are not forming in
the dense gas. In Arp~220 the CN emission from the western nucleus is strong and an indicator
that a fair fraction of the dense gas is in fact in a PDR state. The UV radiation is strongly
affecting the dense molecular clouds here.  Clearly we need more information on the properties of
the dense gas there to find an explanation for the lack of [C II] emission.

\subsection{Line ratios and starburst evolution}

We speculate whether line ratios of dense gas tracers can be used to explore the
evolutionary stage of a starburst. The observed galaxies can be sorted into rough categories
based on their line ratios.

\subsubsection{Warm dense gas --- a young starburst or shocks?}

Objects that show bright HCN emission, but little or no HNC or CN, may be dominated by warm
dense gas early in their starburst development. The Orion KL region is an example of a Galactic
warm ($T \gapprox$ 50 K), dense core where $I$(HCN) is significantly greater than both $I$(CN)
and $I$(HNC)
(e.g., Ungerechts \etal\ 1997). Also the emission from the 10-9 transition of HC$_3$N is brighter
than the CN and HNC emission which is a typical signature of warm, dense gas.
However, shocked gas may be an important part of the molecular ISM in the center of a
starburst galaxy --- in particular if there is a bar in the center where clouds on
intersecting orbits collide. The interaction between supernova remnants at the surrounding
ISM may also lead to the presence of shocked gas. The effect of the shock is to compress and heat
the gas, which in some respects will make it look like an ISM dominated by warm dense cores.
The major difference is that the shock, partially or fully, destroys the dust grains and thereby
releasing molecules, such as SiO, into the ISM. Therefore, SiO emission is often
used as a tracer of shocked gas (e.g., Martin-Pintado \etal\ 1992).
If the shocked gas is allowed to cool after the shock (which may happen quickly since it
has been compressed) the HNC abundance will increase rapidly (S92).
An example of a galaxy that could have a shock dominated ISM is NGC~6240. 
Very strong IR emission is emerging from shock-excited H$_2$ (van der Werf 2001) in between the two
merger nuclei and both HNC and CN line emission is faint relative to that of HCN.
In a high pressure environment the gas may be dense and warm --- but perhaps heated by dissipation
of turbulence rather than very young embedded stars. This may not occur during the very early
stages of star formation, but rather be a form of aftermath.

\subsubsection{PDRs}

For those galaxies where $I$(CN) is $\gapprox$ $I$(HCN) a significant part of the ISM should be
in PDRs. In the Galaxy they are often found in interface regions between HII regions
and molecular clouds (e.g., Jansen \etal\ 1995) and near planetary nebulae. 
CN luminous galaxies are very clearly in the phase where the ISM is being strongly affected
by an intense UV field. The filling factor of UV illuminated gas must be high, and the
clouds probably not too large since the ${[{\rm CN}] \over [{\rm HCN}]}$ abundance ratio drops
dramatically with increasing $A_{\rm V}$. The mechanical impact of the starburst (superwinds - supernovae) may
help in fragmenting the molecular clouds.
Emission from complex molecules, such as HC$_3$N, is faint
because of photodestruction. Since the chemistry now, to a large degree,
involves ion-neutral reactions (see Sect. 4.1) the HNC abundance becomes less dependent
on temperature and $I$(HNC) may be significant even from a warm PDR. For example, the temperature around
a planetary nebula may become very high ($\approx 100$ K) but the HNC abundance is substantial enough
to result in ${{\rm HCN} \over {\rm HNC}}$ line intensity ratios close to unity
(e.g., Herpin \& Cernicharo 2000).

For Arp~220, it appears that HC$_3$N is mainly emerging from the eastern nucleus (see Sect 3.2.1)
which would support the notion of an evolutionary difference between the two nuclei.
Rodriguez-Franco \etal\
(1998) show that the emission from HC$_3$N is bright toward hot, dense cores, while the
HC$_3$N/CN abundance ratio is only $10^{-3}$ in PDRs. Thus the eastern nucleus seems to be in
an earlier evolutionary phase where star formation has just begun. In the mid-IR the western
nucleus is more prominent than the eastern one (Soifer \etal\ 1999).

As discussed in Sect. 4.2.1. the absence of PDR tracers, such as CN emission and the 158 $\mu$m [C II] fine
structure line, does not necessarily mean that the burst is young (or shock dominated). In high-pressure,
dusty ULIRGs it is possible that the formation of PDRs is suppressed --- or that the PDRs are associated with
the diffuse lower density molecular material where the classic PDR lines will not be excited.

\subsubsection{An evolved burst}

Two galaxies (NGC~2623 and IC~694) show fairly bright HNC emission but with undetected CN emission.
Very cold (10 K) and dense gas would result in $I$(HCN)= $0.3-0.5 \times I$(HNC) and 
$I({\rm CN}) < I({\rm HCN})$. In the Galaxy such conditions dominate clouds like TMC-1 and TMC-2
(e.g., Churchwell \etal\  1984).
We know, however, from other studies of IC~694 that the dense gas is warm ($T_{\rm kin} \gapprox 80 K$)
(Aalto \etal\ 1998) and the compact CO nucleus of NGC~2623 likely harbours an ISM similar to that of
IC~694. The starburst may have evolved beyond a strong radiative impact from the stars while the 
chemisty is still dominated by ion-neutral reactions at moderate gas density and relatively high
electron abundance.

\section{Summary}

We have undertaken a SEST and OSO survey of CN and HNC line emission in
a sample of 13 luminous IR galaxies, plus one more ``normal'' starburst (NGC~1808). 
This survey is the first in its kind for IR luminous galaxies.
The main conclusions we draw from this survey are as follows:

\begin{enumerate}

\item We have detected HNC 1-0 in 9 of the galaxies, while CN 1-0
is detected in 10 galaxies. The CN 2-1 lines ($J$=5/2-3/2, $F$=7/2-5/2; J=3/2-1/2 F=5/2-3/2)
(which could only be measured at SEST) are detected in 7 galaxies --- in one (NGC~34)
the CN 1-0 line was not detected, while the 2-1 line was.

\item The line intensitites vary significantly from galaxy to galaxy and the
line ratios relative to the ``standard'' high density tracer, HCN, are far
from constant. The ${{\rm HCN} \over {\rm HNC}}$ ratios vary from 1 to $\gapprox 6$
and the ${{\rm HCN} \over {\rm CN}}$ ratios range from 0.5 to $\gapprox 6$. 
From this we can learn that the actual properties of the dense gas vary 
significantly from galaxy to galaxy, even if their HCN luminosities are similar.

\item We report the detection of HC$_3$N 10-9 emission in Arp~220
at 40\% of the HNC 1-0 intensity. In NGC~1808 the line is tentatively detected.

\item We find that the IR luminous galaxies are at least as HNC luminous as more
nearby (IR-fainter) galaxies. We conclude that the HNC emission is not a reliable
tracer of cold (10 K) gas in the center
of luminous IR galaxies, the way it may be in clouds in the disk of the Milky Way.
Instead, we list several alternative (testable) explanations. Provided that the situation in the
centers of starburst galaxies can be approximated with a steady state chemistry, the
observed ${{\rm HCN} \over {\rm HNC}}$ line ratios can be explained by the emission
originating in gas of
densities $n \lapprox 10^5$ $\cmmd$ where the chemistry is dominated by ion-neutral,
instead of neutral-neutral, reactions. The temperature dependence of the HNC production
in the ion-neutral reactions is strongly suppressed.

\item We were surprised to find that the CN emission of the more luminous galaxies is
often faint. This can be consistent with the relative faintness of other PDR tracers
such as the 158 $\mu$m [C II] line. Both these phenomena can be understood in terms of
a two-component molecular ISM consisting of low density molecular gas surrounding dense
clouds. However, the relative brightness of CN in Arp~220 may be difficult to reconcile
with its relative [C II] faintness.  

\item Within Arp~299 (the merging pair IC~694 and NGC3690) there is a line ratio difference
between the two nuclei. NGC~3690 has a CN luminosity twice that of HCN and its ISM is thus
strongly affected by UV radiation while CN is not detected towards IC~694. This reflects an
evolutionary gradient in the burst --- or that the star formation activity takes place in
very different environments in the two galaxies.
For Arp~220 there may also be an evolutionary gradient between the two nuclei because of
differences in emission velocities for CN and HC$_3$N. 

\item The measured line ratios can in general be used to identify stages in the starburst
evolution, but we conclude that the dichotomy between scenarios is a complication. For
example, faint HNC emission is expected both in a shock dominated ISM as well as for a
cloud ensemble dominated by dense warm gas in the very early stages of a starburst.

\end{enumerate}

\acknowledgements{Many thanks to the OSO and SEST staff for their help. We are grateful
to M. Walmsley and P. Schilke for discussions on the HNC chemistry. We thank the referee, F.
Herpin, for many useful comments and suggestions.}

\end{document}